\documentclass{article}

\usepackage[nonatbib,final]{neurips_2022}

\usepackage[utf8]{inputenc} % allow utf-8 input
\usepackage[T1]{fontenc}    % use 8-bit T1 fonts
\usepackage{hyperref}       % hyperlinks
\usepackage{url}            % simple URL typesetting
\usepackage{booktabs}       % professional-quality tables
\usepackage{amsfonts}       % blackboard math symbols
\usepackage{nicefrac}       % compact symbols for 1/2, etc.
\usepackage{microtype}      % microtypography
\usepackage{xcolor}         % colors

\usepackage[english]{babel}
\usepackage{csquotes}
\usepackage{physics}
\usepackage{mathtools}
\usepackage{isomath}
\usepackage{float}
\usepackage{wrapfig}
\usepackage{multirow}
\usepackage{etoolbox}
\usepackage[backend=biber, style=phys, sorting=none]{biblatex}
\usepackage{subcaption}
\usepackage{chngcntr}

\usepackage{amsthm}

\title{Learning Feynman Diagrams using Graph Neural Networks}

\author{%
  Harrison Mitchell\\
  Cavendish Laboratory\\
  University of Cambridge\\
  \texttt{hgwm2@cantab.ac.uk} \\
  \And 
  Alexander Norcliffe\\
  Computer Laboratory\\
  University of Cambridge\\
  \texttt{alin2@cam.ac.uk}
  \And
  Pietro Li\`{o}\\
  Computer Laboratory\\
  University of Cambridge\\
  \texttt{pl219@cam.ac.uk}
}

\begin{document}

\maketitle

\begin{abstract}
In the wake of the growing popularity of machine learning in particle physics, this work finds a new application of geometric deep learning on Feynman diagrams to make accurate and fast matrix element predictions with the potential to be used in analysis of quantum field theory. This research uses the graph attention layer which makes matrix element predictions to 1 significant figure accuracy above 90\% of the time. Peak performance was achieved in making predictions to 3 significant figure accuracy over 10\% of the time with less than 200 epochs of training, serving as a proof of concept on which future works can build upon for better performance. Finally, a procedure is suggested, to use the network to make advancements in quantum field theory by constructing Feynman diagrams with effective particles that represent non-perturbative calculations.

\end{abstract}

\section{Introduction and Related Work}
Particle physics has recently seen a rise in the popularity of machine learning \cite{feickert}, \cite{guest}, \cite{shlomi}. Applications include, analysing particle jets \cite{barnard} \cite{de_Oliveira}, tracking particle paths \cite{duarte} and detecting collider events \cite{alanazi}. This work explores a new application of geometric deep learning \cite{scarselli}\cite{bronstein} on graph neural networks (GNNs) by predicting matrix elements from Feynman diagrams for simple particle interactions. Analytically, matrix elements are computed from Feynman diagrams, a pictorial representation of a particle interaction, using a set of rules called the Feynman rules, which translate the diagram into a mathematical expression. The matrix elements then find wide use in computing probabilities in particle physics. Graph neural networks are a type of network that operate on graph valued data, to make node, edge or graph level predictions. In this work, a graph level prediction is used for matrix elements from a graph representation of Feynman diagrams. The convolutional layer used is the graph attention layer (GAT) \cite{velickovic}, which previously has shown application in text classification \cite{yao}, material property predictions \cite{louis} and classical physics \cite{wang}. This work also aims to extend the testing of the performance of the GAT layer and the effectiveness of its attention head mechanism on Feynman diagrams. The advantage of using machine learning for physics is partly in faster computation time for complicated matrix elements, but the envisioned future of this study is that the neural network learns all Feynman rules so that it can be connected to a dynamic graph network (for example, the temporal graph network \cite{rossi}). Such a network would progressively construct new nodes and edges, creating new Feynman diagrams which would be passed to the learned GNN whose output would be used to fit to matrix elements from experimental data. This may predict effective Feynman diagrams in non-perturbative regimes, such as low energy QCD, where truncating the infinite series of perturbations does not approximate the matrix elements. Many of the matrix elements that were used extensively in the learning process for the neural networks were adapted from the 1995 work by Borodulin \cite{borodulin}.

\section{Method}

\begin{figure}[t]
    \centering
    \begin{subfigure}[h]{0.4\textwidth}
        \includegraphics[width=\textwidth]{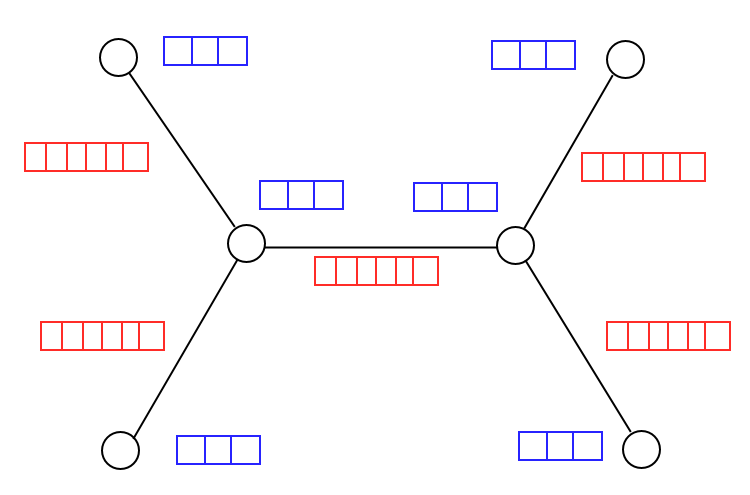}
        \caption{An example tree-level graph, that the network can take as an input. The work of this paper is on diagrams with this structure but with different edge features. Blue vectors are node features - interactions. Red vectors are edge features - particles.}
        \label{fig: graph enc}
    \end{subfigure}
    \hfill
    \begin{subfigure}[h]{0.5\textwidth}
        \includegraphics[width=\textwidth]{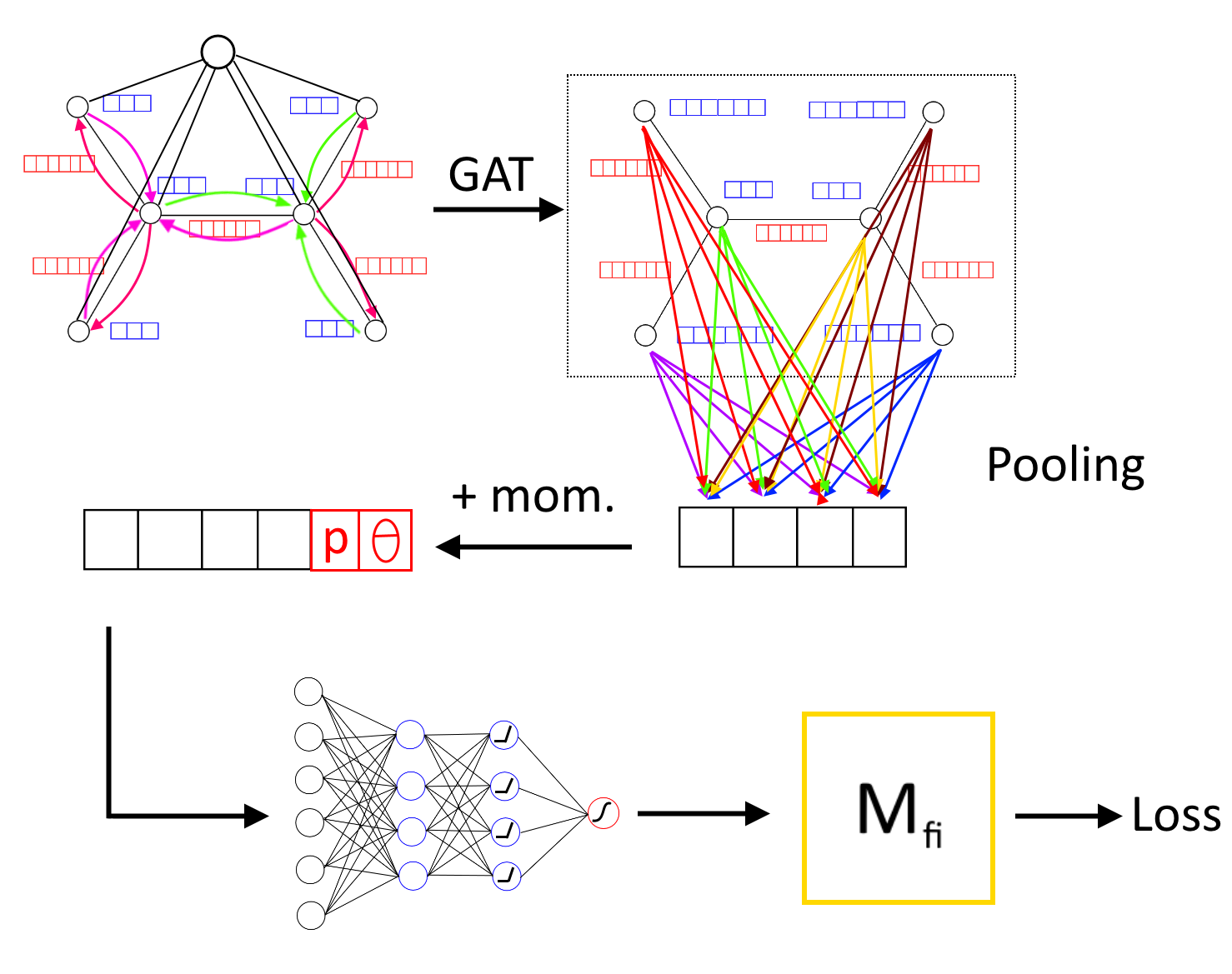}
        \caption{Top left image depicts the action of the GAT layer. Top right is a portrayal of the output of the GAT layers being pooled into a graph representation and momenta added. This is passed to an FCN to get the predictions.}
        \label{fig: GNN architecture}
    \end{subfigure}\caption{Sample input (left) and network architecture (right).}
\end{figure}

\subsection{Dataset encoding}
Feynman diagrams have a natural graph representation as seen in Figure \ref{fig: graph enc}, with interaction vertices being represented by nodes, and particles being represented by edges. The edge features contain all the information to describe a particle, whilst the targets are the matrix elements, averaged over the helicity combinations. In our graph encodings, the momenta are not included in the edge features and are instead concatenated after the GNN has produced a graph representation. The concatenation process is outlined in Figure \ref{fig: GNN architecture}. Since the Feynman rules are independent of momentum in both QED and QCD, the representation of each graph should be as well. This method provides scalability by removing the need for multiple diagrams for each possible helicity combination. This is a more efficient procedure when the number of Feynman diagrams is large. The particle encoding is given in Equation \ref{eqn: edge feat}, where masses are on-shell and in mega electronvolts (MeV).

\begin{equation}\label{eqn: edge feat}
    \text{edge vector} = [m,S,I_{L}^3,Y_L,I_{R}^3,Y_R,r,g,b,\bar{r},\bar{g},\bar{b}]
\end{equation}

All data is given in the centre of mass frame with azimuthal angle \(\phi=0\) and particles aligned with the \(z\)-axis, so that only the polar angle, \(\theta\), and the momentum, \(p\), are needed to categorise the interaction. Each data point includes a graph, representing the Feynman diagram, and a \((p,\theta)\) pair. A large dataset consisting of 80,000 such data points with ultra-relativistic momenta in the range \([1, 1000]\)GeV, was then constructed and training was carried out on subsets containing only the appropriate Feynman diagrams for the experiment. Each subset consisted of 2000 data points, sampled evenly over the range of target matrix elements. The network is fed three random splits of the dataset: training, validation and test datasets The split ratio was 6:2:2 respectively. The training dataset is what the model sees and uses to conduct back-propagation and the order is shuffled each epoch. The validation dataset is passed through the model at the end of each epoch. The test dataset is passed through after training.

A global node was also added to improve the training speed and performance of the network. The three interaction strengths for QED, QCD and the weak force (\(\alpha_{QED},\alpha_S,\alpha_W\)) were stored in the global node vector, each given at zero energy. The logarithmic running of these constants can be learnt during propagation through subsequent node embeddings of the global node in the forward pass. This addition reduced the number of required layers and trainable parameters in the GNN since node information on opposite ends of the graph only needs two layers to pass to each other via the global node. It was seen that even when the number of attention heads and layers in the GNN were reduced, the network performed similarly to before adding the global node.

\subsection{Network Architecture}

The aim is that the GNN will create a representation that encodes information for the Feynman rules. The network consists of a GNN connected to a fully connected network (FCN). The GNN architecture is built using three key components: a convolutional layer, a linear layer and a pooling layer. The convolutional layer used is the graph attention layer \cite{velickovic}. The linear layer invokes simple fully connected layer with a leaky ReLu activation function, needed to condense the size of the graph, since the attention mechanism generates a larger graph depending on the number of attention heads. The pooling layer uses global sum pooling in order to capture information across all nodes as well as distinguishing different sized graphs by the magnitude of the values in the representation. The graph representation is then passed to the FCN consisting of two hidden layers and a leaky ReLU activation function. The final output layer has just one neuron whose value is passed through a sigmoid activation function so that the output is squeezed between zero and one. The architecture is summarised in the schematic in Figure \ref{fig: GNN architecture}. Further implementation details are included in Appendix \ref{apdx: imple}.

\section{Results and Discussion}\label{chp: res and disc}

The network was trained on high energy tree-level QED diagrams involving electrons and muons on which the model accurately predicts matrix elements for several processes. There were 2 main metrics used to evaluate performance: \(L^1\) loss and the accuracy. Accuracy is defined as the proportion of results in the dataset that are the same as the target after rounding up to a number of decimal places (d.p.), \(a\), as shown in Equation \ref{eqn: acc} .

\begin{equation}\label{eqn: acc}
   \mathcal{A}(\{y_i\},\{\hat{y}_i\},a)=\frac{1}{N}\sum_i^N \left(1-\Theta\left[L^1\left(\lceil y_i\rceil_a,\lceil\hat{y}_i\rceil_a \right)\right]\right)
\end{equation}

Where \(\lceil x \rceil_a\) means round up to the \(a^{th}\) decimal place and \(\Theta(x)\) is the step function. The predicted matrix elements are plotted against the polar angle in the range \([0,2\pi]\). At large momenta, the matrix element is independent of momentum, and so the variation is in the angle. These metrics are tabulated in each experiment's data tables. A further extension of the model is tested on QCD, involving quark diagrams. The results are included in Appendix \ref{apdx: QCD}.

\subsection{Three diagram dataset}

\begin{figure}[t]
    \centering
    \begin{subfigure}[h]{0.45\textwidth}
        \includegraphics[width=\textwidth]{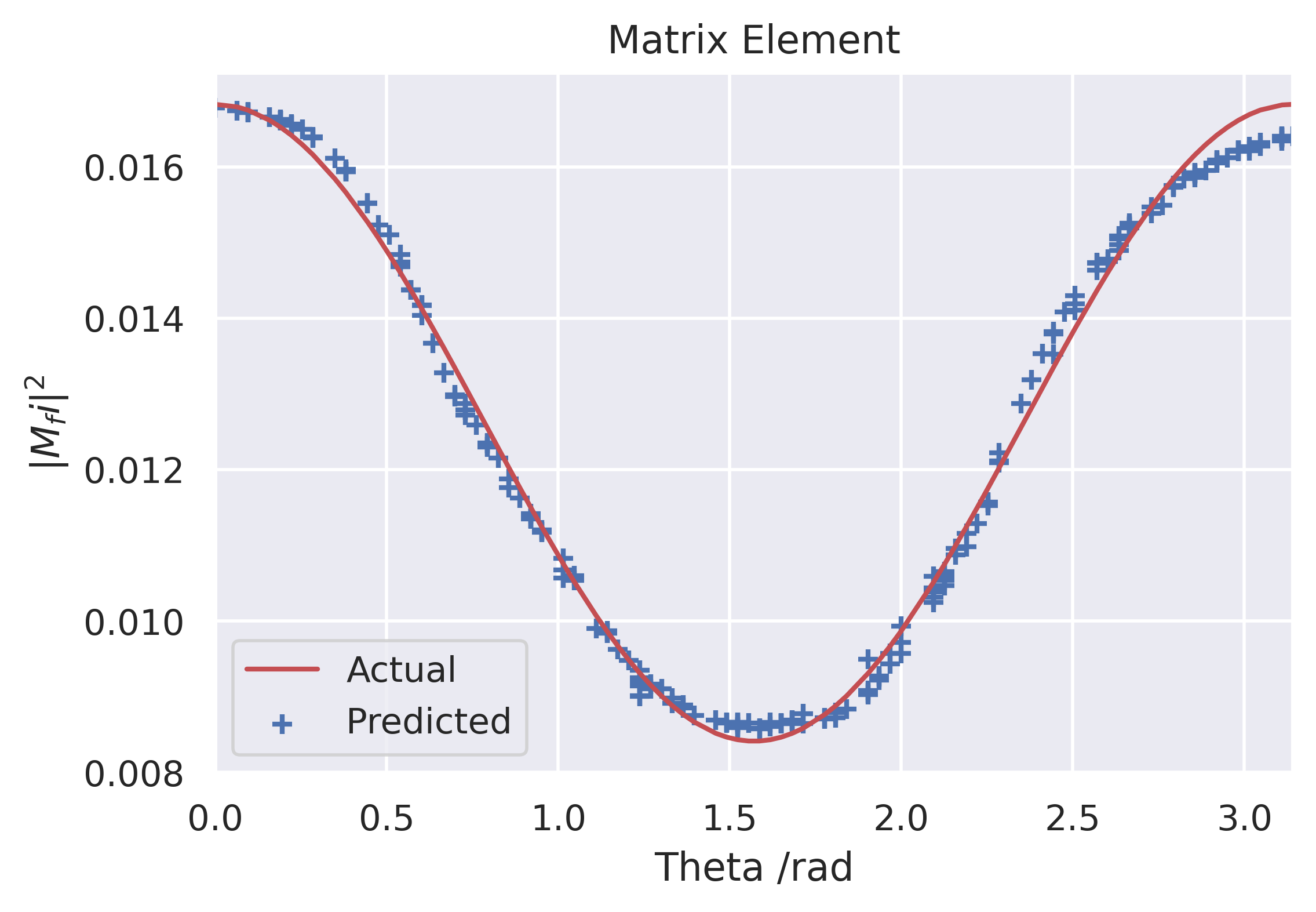}
        \caption{Results for \(e^-e^+\to\mu^-\mu^+\) from a GNN that has seen both matrix elements}
        \label{fig: mu 3}
    \end{subfigure}
    \hfill
    \begin{subfigure}[h]{0.45\textwidth}
        \includegraphics[width=\textwidth]{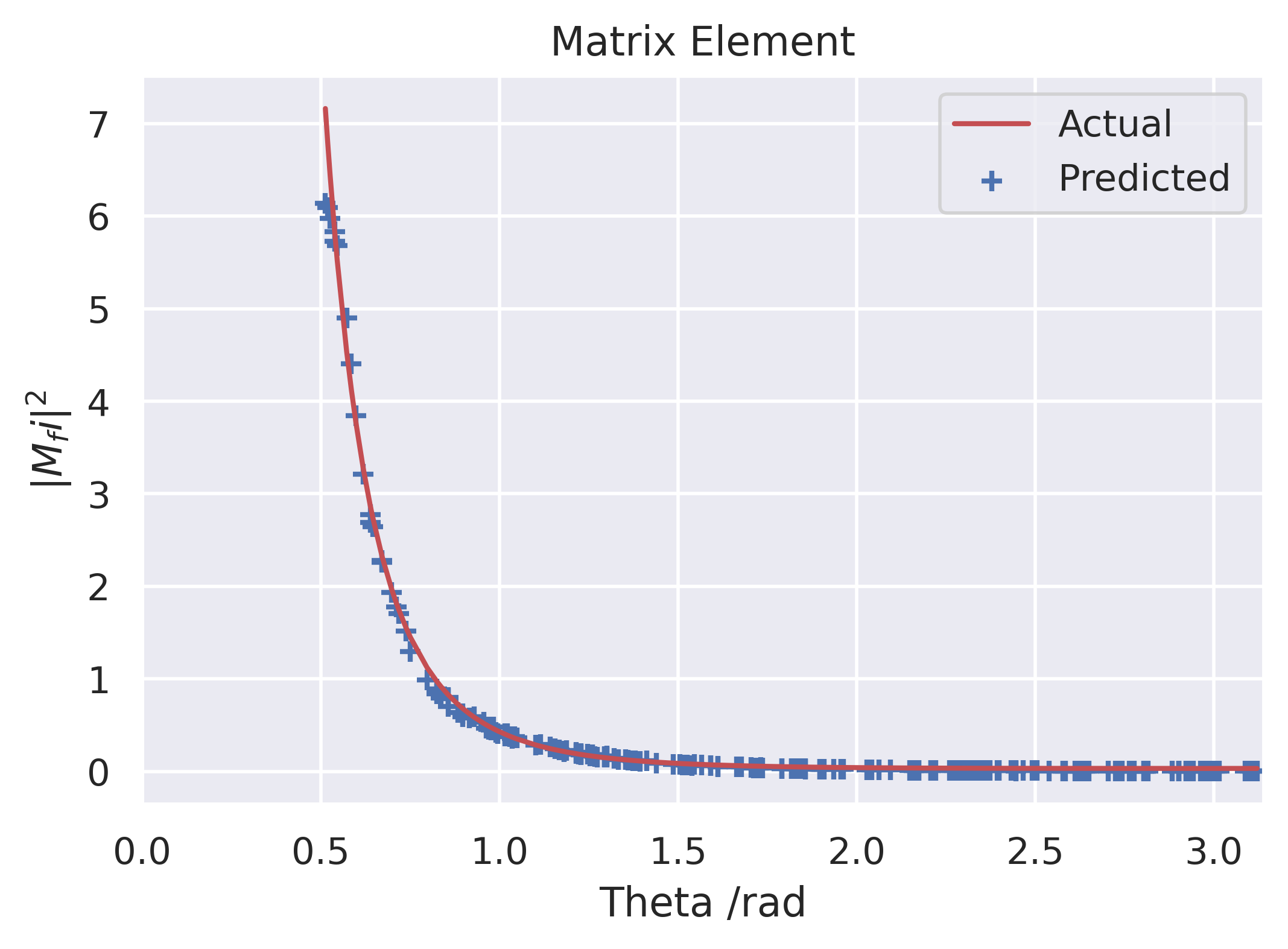}
        \caption{Results for \(e^-e^+\to e^-e^+\) Bhabha scattering matrix elements.}
        \label{fig: e 3}
    \end{subfigure}\caption{Results for training on three diagrams}
\end{figure}

The initial model was tested on two graphs consisting of three Feynman diagrams: the electron-positron to muon-antimuon s-channel diagram; and the two Bhabha electron-positron to electron-positron s and t-channel diagrams, batched into one graph with ultra-relativistic matrix element. %Might need to include some diagrams and equations
Since both graphs are combined into a single dataset the GNN is forced to distinguish between the two different inputs, testing its ability to construct a sufficiently detailed graph representation to feed to the FCN to communicate which matrix element is to be fitted.

Figures \ref{fig: mu 3} and \ref{fig: e 3} exhibit the model's performance in fitting over the range of angles. The predictions are good with a small discrepancy for the muon matrix element at large angles. More detail on the loss and accuracy values are given in Table \ref{tab: 3 graphs}, where it can be seen that the accuracy to one decimal place is high but just shy of 100\% because the model predicts some matrix elements very close to the target, but some much farther away so that the \(L^1\) loss is consistently low but allowing for a much tighter fit for some of the data points reaching  two and three decimal place accuracy over 81\% and 11\% of the time, respectively.

\begin{table}[h]
  \caption{Loss and accuracy values for the three combined Feynman diagrams for \(e^-e^+\to\mu^-\mu^+\) and \(e^-e^+\to e^-e^+\)}
  \label{tab: 3 graphs}
  \centering
  \begin{tabular}{cc}
    \toprule
    Test metric     & Data      \\
    \midrule
    Test accuracy (1d.p.) & 99.00\%\\
    Test accuracy (2d.p.) & 81.50\% \\
    Test accuracy (3d.p.) & 11.50\% \\
    Test L1 Loss & 0.0049\\
    \bottomrule
  \end{tabular}
\end{table}

\subsection{Tree-level QED dataset for electrons and muons}

\begin{figure}[t]
    \centering
    \begin{subfigure}[t]{0.45\textwidth}
        \includegraphics[width=\textwidth]{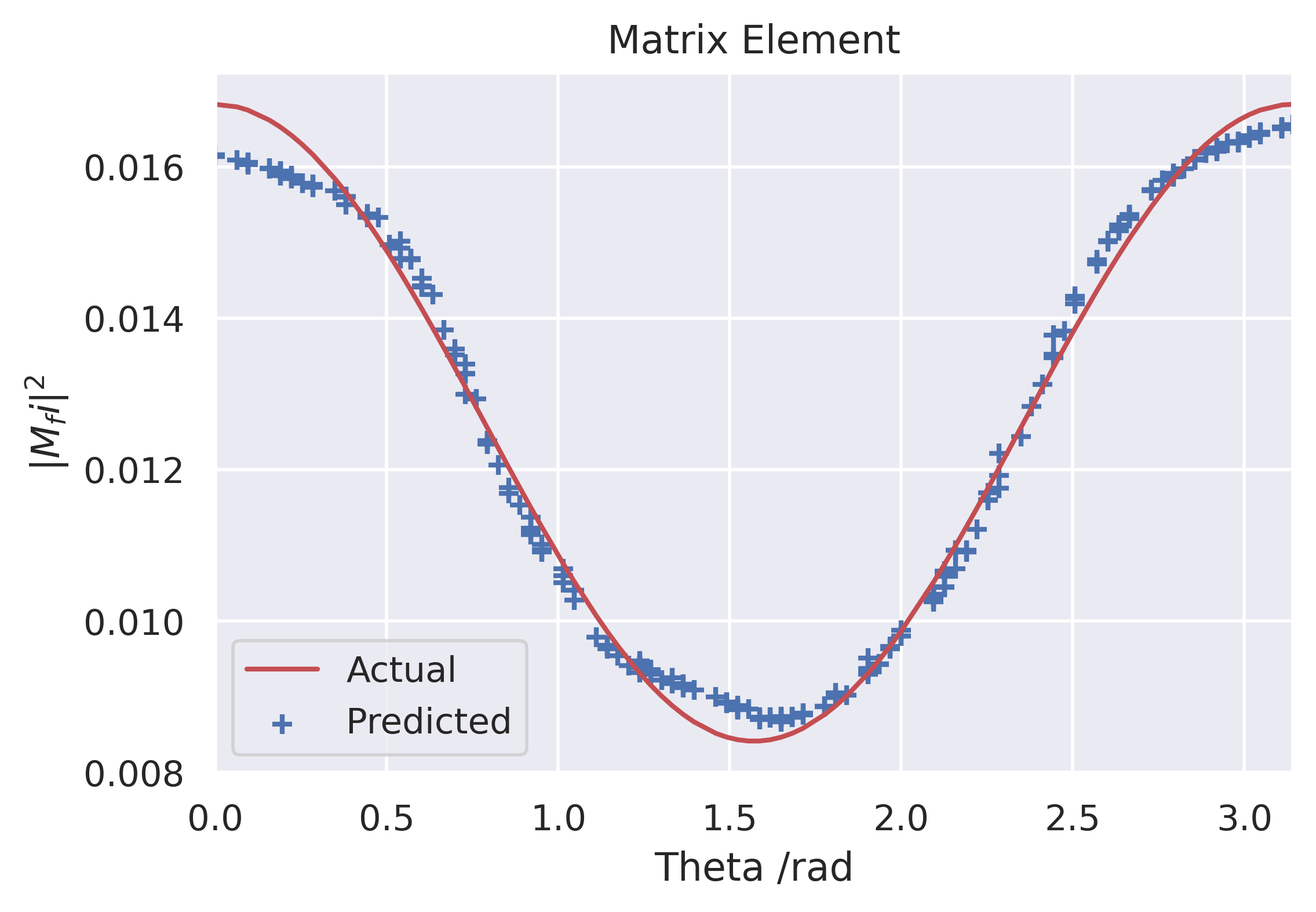}
        \caption{Results from the fully trained network on all tree-level QED with electrons and muons, for the \(e^-e^+\to\mu^-\mu^+\) matrix element.}
        \label{fig: mu full}
    \end{subfigure}
    \hfill
    \begin{subfigure}[t]{0.45\textwidth}
        \includegraphics[width=\textwidth]{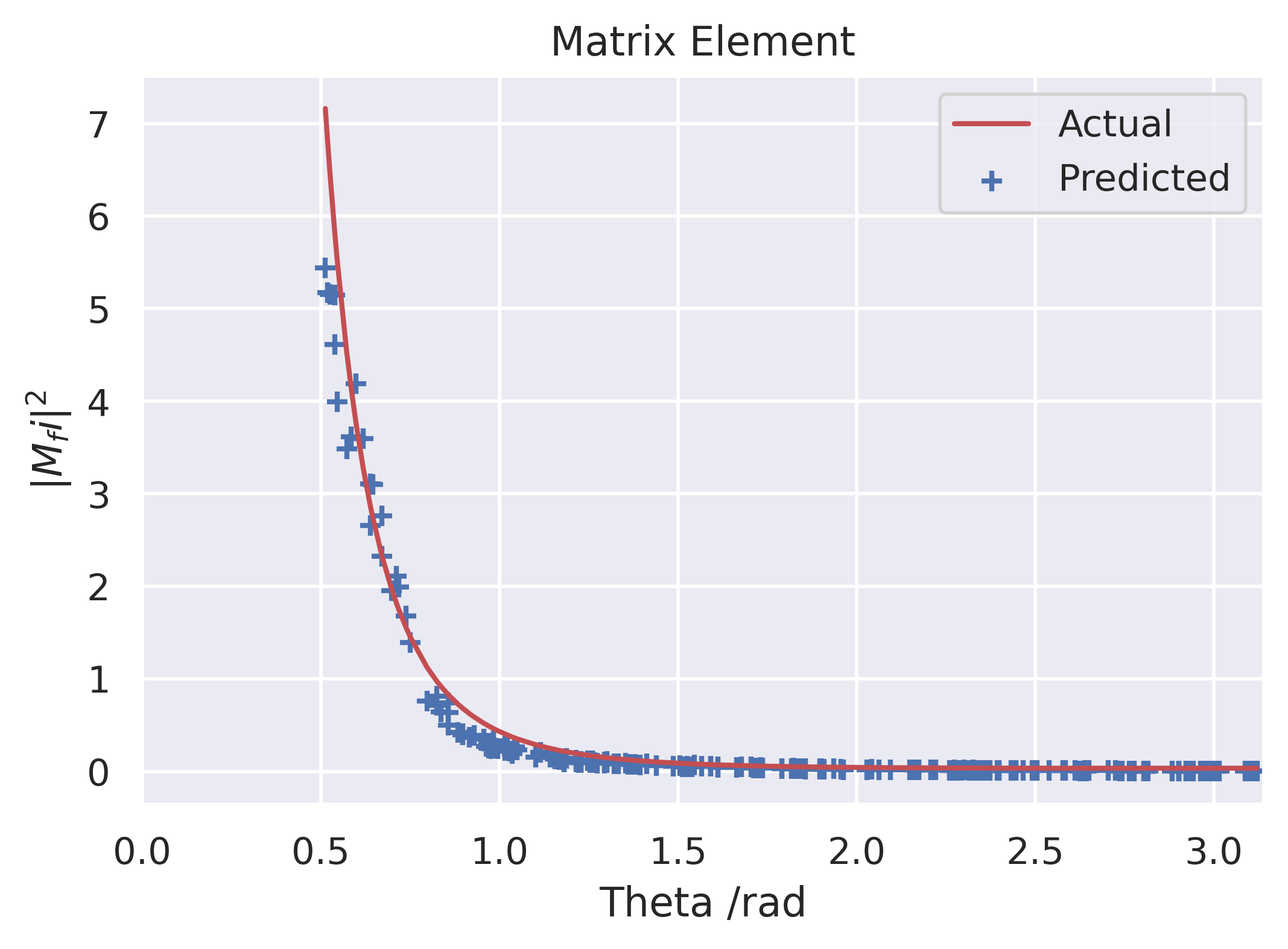}
        \caption{Results from the fully trained network, for the \(e^-e^+\to e^-e^+\) matrix element.}
        \label{fig: e full}
    \end{subfigure}\caption{Training on full dataset}
\end{figure}

The dataset was expanded to encompass six graphs with four high energy matrix elements at tree-level QED, for each of the permutations of electrons and muons in the initial and final states and their allowed s and t-channels. There is a drop in performance but the overall target shape is maintained. Figure \ref{fig: mu full} shows the predicted and theoretical matrix elements for the same s-channel \(e^-e^+\to\mu^-\mu^+\) interaction. This time, this is one of many graphs that the network is able to predict, but despite the increase in number of points to fit, we still see strong performance. For this interaction the discrepancy at the extreme angles 0 and \(2\pi\) is more noticeable, but there is additional error at the minimum \(\theta=\pi\). Since the model performs best in the regions that are close to linear, this may be indicative of a requirement for additional non-linearities.

Figure \ref{fig: e full} shows the predicted and theoretical matrix element for the s and t-channel \(e^-e^+\to e^-e^+\) interactions, where we see much stronger agreement. Increasing the number of dense neurons showed signs of improvement, but significantly increased the training time, but since this target consists of two largely linear regimes, the models performs well considering the training time. Table \ref{tab: full QED} shows the metrics for the learned GNN. The most noticeable difference is the much higher loss. This behaviour is expected for a larger dataset as there are more points for the network to fit to, but despite increasing the width of the FCN, the GNN was unable to fit as accurately as before. This may have been, in part, due to a need for an increased number of neurons in the FCN to be able to decode and fit multiple different functions depending on the graph representation it sees. Alternatively it may be that the GNN's graph representations were not large enough to discriminate between the graphs. The graph representation was kept to a constant dimension in the hope that the strong fit to the smaller three diagram dataset implied that the Feynman rules were sufficiently encoded. Increasing this may help to encode further detail of the Feynman rules with different starting and final states. Finally, the minima search may be further improved by smoothing the landscape with stochastic weight averaging by allowing the network to find lower minima and therefore reduce the loss.

However, the ability to identify different graphs and reconstruct different functions suggests that the GNN has the capacity to encode the Feynman rules into its graph representation. With some improvement in implementation it is possible that the GNN can make more accurate predictions and for a wider range of particle states.

\begin{table}[h]
    \caption{Loss and accuracy values for all tree-level QED Feynman diagrams involving all permutations of \(e^-e^+\) and \(\mu^-\mu^+\) pairs}
    \label{tab: full QED}
    \centering
    \begin{tabular}{cc}
        \toprule
        Test metric & Data \\
        \midrule
        Test accuracy (1d.p.) & 96.75\%\\
        Test accuracy (2d.p.) & 44.00\% \\
        Test accuracy (3d.p.) & 0.75\% \\
        Test L1 Loss & 0.015\\
        \bottomrule
    \end{tabular}
\end{table}

%%%%%%%%%%%%%%%%%%%%%%
%%%%% Conclusion %%%%%
%%%%%%%%%%%%%%%%%%%%%%
\section{Conclusion}

Graph neural networks have been shown to be able to make basic matrix element predictions by fitting to analytical data. This provides improved computation time for matrix element calculations and scales well when introducing new matrix elements. This work suggests that it is possible for the GNN's graph representation to encode the elements of the Feynman rules, and for the FCN to decode it into a matrix element. Following on from the research in the field of geometric deep learning, graph attention networks have been shown to be effective in regression problems for Feynman diagrams. This supports the use of geometric deep learning in particle physics beyond experimental analysis. The aspiration is that this work will be extended to have learnt a full dataset with all fundamental forces that, in itself, would be useful in making experimental predictions. To further build on this, if the dataset is extended to loop diagrams, the network will learn its own procedure to deal with renormalization. This could then be connected to a network for dynamic graphs, which will learn the edges and nodes to build, which upon passing to the GNN, correctly fit to experimental data. The constructed diagrams are then ideally interpretable as Feynman diagrams, corresponding to physical processes, and if not, then at least it may provide effective Feynman diagrams, whose matrix elements represent an infinite series in non-perturbative QFT.

\newpage

\begin{ack}
We thank Bentley Carr for his time and insight during the research of this project. Additionally we would like to take the chance to thank the reviewers for raising challenging points that drove us to find new perspectives and better the future direction of this work.

Finally, we also thank the Cavendish Laboratory and Computer Laboratory at the University of Cambridge for facilitating the original research.
\end{ack}

\nocite{*}
\printbibliography

@article{borodulin,
  doi = {10.48550/ARXIV.HEP-PH/9507456},
  url = {https://arxiv.org/abs/hep-ph/9507456},
  author = {Borodulin, V. I. and Rogalyov, R. N. and Slabospitsky, S. R.},
  title = {CORE 2.1 (COmpendium of RElations, Version 2.1)},
  publisher = {arXiv},
  year = {1995},
  copyright = {Assumed arXiv.org perpetual, non-exclusive license to distribute this article for submissions made before January 2004}
}

@article{bronstein,
	doi = {10.1109/msp.2017.2693418},
	url = {https://doi.org/10.1109%2Fmsp.2017.2693418},
	year = 2017,
	month = {7},
	publisher = {Institute of Electrical and Electronics Engineers ({IEEE})},
	volume = {34},
	number = {4},
	pages = {18--42},
	author = {Michael Bronstein and Joan Bruna and Yann LeCun and Arthur Szlam and Pierre Vandergheynst},
	title = {Geometric Deep Learning: Going beyond Euclidean data},
	journal = {{IEEE} Signal Processing Magazine}
}

@article{scarselli,
    author={Scarselli, Franco and Gori, Marco and Tsoi, Ah Chung and Hagenbuchner, Markus and Monfardini, Gabriele},
    journal={IEEE Transactions on Neural Networks},
    title={The Graph Neural Network Model},
    year={2009},
    volume={20},
    number={1},
    pages={61-80},
    doi={10.1109/TNN.2008.2005605}
}

@article{velickovic,
  doi = {10.48550/ARXIV.1710.10903},
  url = {https://arxiv.org/abs/1710.10903},
  author = {Velickovic, Petar and Cucurull, Guillem and Casanova, Arantxa and Romero, Adriana and Liò, Pietro and Bengio, Yoshua},
  title = {Graph Attention Networks},
  publisher = {arXiv},
  year = {2017},
  copyright = {arXiv.org perpetual, non-exclusive license}
}

@article{rossi,
  doi = {10.48550/ARXIV.2006.10637},
  url = {https://arxiv.org/abs/2006.10637},
  author = {Rossi, Emanuele and Chamberlain, Ben and Frasca, Fabrizio and Eynard, Davide and Monti, Federico and Bronstein, Michael},
  title = {Temporal Graph Networks for Deep Learning on Dynamic Graphs},
  publisher = {arXiv},
  year = {2020},
  copyright = {arXiv.org perpetual, non-exclusive license}
}

@article{bruna,
  doi = {10.48550/ARXIV.1312.6203},
  url = {https://arxiv.org/abs/1312.6203},
  author = {Bruna, Joan and Zaremba, Wojciech and Szlam, Arthur and LeCun, Yann},
  title = {Spectral Networks and Locally Connected Networks on Graphs},
  publisher = {arXiv},
  year = {2013},
  copyright = {arXiv.org perpetual, non-exclusive license}
}

@article{mertig_G1991,
title = {Feyn Calc - Computer-algebraic calculation of Feynman amplitudes},
journal = {Computer Physics Communications},
volume = {64},
number = {3},
pages = {345-359},
year = {1991},
issn = {0010-4655},
doi = {https://doi.org/10.1016/0010-4655(91)90130-D},
url = {https://www.sciencedirect.com/science/article/pii/001046559190130D},
author = {R. Mertig and M. Böhm and A. Denner}
}

@article{Shtabovenko_2016,
	doi = {10.1016/j.cpc.2016.06.008},
	url = {https://doi.org/10.1016%2Fj.cpc.2016.06.008},
	year = 2016,
	month = {10},
	publisher = {Elsevier {BV}},
	volume = {207},
	pages = {432--444},
	author = {Vladyslav Shtabovenko and Rolf Mertig and Frederik Orellana},
	title = {New developments in {FeynCalc} 9.0},
	journal = {Computer Physics Communications}
}

@article{Shtabovenko_2020,
	doi = {10.1016/j.cpc.2020.107478},
	url = {https://doi.org/10.1016%2Fj.cpc.2020.107478},
	year = 2020,
	month = {11},
	publisher = {Elsevier {BV}},
	volume = {256},
	pages = {107478},
	author = {Vladyslav Shtabovenko and Rolf Mertig and Frederik Orellana},
	title = {{FeynCalc} 9.3: New features and improvements},
	journal = {Computer Physics Communications}
}

@article{hamilton_2017,
  doi = {10.48550/ARXIV.1706.02216},
  url = {https://arxiv.org/abs/1706.02216},
  author = {Hamilton, William L. and Ying, Rex and Leskovec, Jure},
  title = {Inductive Representation Learning on Large Graphs},
  publisher = {arXiv},
  year = {2017},
  copyright = {arXiv.org perpetual, non-exclusive license}
}

@article{hamilton_2020,
    author={Hamilton, William L.},
    title={Graph Representation Learning},
    journal={Synthesis Lectures on Artificial Intelligence and Machine Learning},
    url = {https://www.cs.mcgill.ca/~wlh/grl_book/},
    volume={14},
    number={3},
    pages={1-159},
    publisher={Morgan and Claypool}
}

@article{gilmer,
  doi = {10.48550/ARXIV.1704.01212},
  url = {https://arxiv.org/abs/1704.01212},
  author = {Gilmer, Justin and Schoenholz, Samuel S. and Riley, Patrick F. and Vinyals, Oriol and Dahl, George E.},
  
  title = {Neural Message Passing for Quantum Chemistry},
  
  publisher = {arXiv},
  
  year = {2017},
  
  copyright = {arXiv.org perpetual, non-exclusive license}
}

@article{dai,
  doi = {10.48550/ARXIV.1603.05629},
  url = {https://arxiv.org/abs/1603.05629},
  author = {Dai, Hanjun and Dai, Bo and Song, Le},
  title = {Discriminative Embeddings of Latent Variable Models for Structured Data},
  publisher = {arXiv},
  year = {2016},
  copyright = {arXiv.org perpetual, non-exclusive license}
}

@article{feickert,
  doi = {10.48550/ARXIV.2102.02770},
  
  url = {https://arxiv.org/abs/2102.02770},
  
  author = {Feickert, Matthew and Nachman, Benjamin},
  
  title = {A Living Review of Machine Learning for Particle Physics},
  
  publisher = {arXiv},
  
  year = {2021},
  
  copyright = {Creative Commons Attribution 4.0 International}
}

@article{barnard,
	doi = {10.1103/physrevd.95.014018},
  
	url = {https://doi.org/10.1103%2Fphysrevd.95.014018},
  
	year = 2017,
	month = {1},
  
	publisher = {American Physical Society ({APS})},
  
	volume = {95},
  
	number = {1},
  
	author = {James Barnard and Edmund Noel Dawe and Matthew J. Dolan and Nina Rajcic},
  
	title = {Parton shower uncertainties in jet substructure analyses with deep neural networks},
  
	journal = {Physical Review D}
}

@article{de_Oliveira,
	doi = {10.1007/jhep07(2016)069},
  
	url = {https://doi.org/10.1007%2Fjhep07%282016%29069},
  
	year = 2016,
	month = {7},
  
	publisher = {Springer Science and Business Media {LLC}
},
  
	volume = {2016},
  
	number = {7},
  
	author = {Luke de Oliveira and Michael Kagan and Lester Mackey and Benjamin Nachman and Ariel Schwartzman},
  
	title = {Jet-images {\textemdash} deep learning edition},
  
	journal = {Journal of High Energy Physics}
}

@article{guest,
author = {Guest, Dan and Cranmer, Kyle and Whiteson, Daniel},
title = {Deep Learning and Its Application to LHC Physics},
journal = {Annual Review of Nuclear and Particle Science},
volume = {68},
number = {1},
pages = {161-181},
year = {2018},
doi = {10.1146/annurev-nucl-101917-021019},

URL = { 
        https://doi.org/10.1146/annurev-nucl-101917-021019
    
},
eprint = { 
        https://doi.org/10.1146/annurev-nucl-101917-021019
    
}
}

@incollection{duarte,
	doi = {10.1142/9789811234033_0012},
	url = {https://doi.org/10.1142%2F9789811234033_0012},
	year = 2022,
	month = {2},
	publisher = {{WORLD} {SCIENTIFIC}},
	pages = {387-436},
	author = {Javier Duarte and Jean-Roch Vlimant},
	title = {Graph Neural Networks for Particle Tracking and Reconstruction},
	booktitle = {Artificial Intelligence for High Energy Physics}
}

@article{shlomi,
	doi = {10.1088/2632-2153/abbf9a},
  
	url = {https://doi.org/10.1088%2F2632-2153%2Fabbf9a},
  
	year = 2021,
	month = {1},
  
	publisher = {{IOP} Publishing},
  
	volume = {2},
  
	number = {2},
  
	pages = {021001},
  
	author = {Jonathan Shlomi and Peter Battaglia and Jean-Roch Vlimant},
  
	title = {Graph neural networks in particle physics},
  
	journal = {Machine Learning: Science and Technology}
}

@inproceedings{alanazi,
	doi = {10.24963/ijcai.2021/588},
  
	url = {https://doi.org/10.24963%2Fijcai.2021%2F588},
  
	year = 2021,
	month = {8},
  
	publisher = {International Joint Conferences on Artificial Intelligence Organization},
  
	author = {Yasir Alanazi and Nobuo Sato and Pawel Ambrozewicz and Astrid Hiller-Blin and Wally Melnitchouk and Marco Battaglieri and Tianbo Liu and Yaohang Li},
  
	title = {A Survey of Machine Learning-Based Physics Event Generation},
  
	booktitle = {Proceedings of the Thirtieth International Joint Conference on Artificial Intelligence}
}

@article{butter,
  doi = {10.48550/ARXIV.2008.08558},
  
  url = {https://arxiv.org/abs/2008.08558},
  
  author = {Butter, Anja and Plehn, Tilman},
  
  title = {Generative Networks for LHC events},
  
  publisher = {arXiv},
  
  year = {2020},
  
  copyright = {arXiv.org perpetual, non-exclusive license}
}

@InProceedings{wang,
  title = 	 {Bridging Physics-based and Data-driven modeling for Learning Dynamical Systems},
  author =       {Wang, Rui and Maddix, Danielle and Faloutsos, Christos and Wang, Yuyang and Yu, Rose},
  booktitle = 	 {Proceedings of the 3rd Conference on Learning for Dynamics and Control},
  pages = 	 {385--398},
  year = 	 {2021},
  editor = 	 {Jadbabaie, Ali and Lygeros, John and Pappas, George J. and Parrilo, Pablo and Recht, Benjamin and Tomlin, Claire J. and Zeilinger, Melanie N.},
  volume = 	 {144},
  series = 	 {Proceedings of Machine Learning Research},
  publisher =    {PMLR},
  url = 	 {https://proceedings.mlr.press/v144/wang21a.html},
}

@article{louis,
    doi = {10.1039/D0CP01474E},
    year = {2020},
    volume = {22},
    title = {Graph convolutional neural networks with global attention for improved materials property prediction},
    pages = {18141-18148},
    author = {Louis, Steph-Yves and Zhao, Yong and Nasiri, Alireza and Wang, Xiran and Song,Yuqi and Liu, Fei and Hu, Jianjun},
    journal = {Phys. Chem. Chem. Phys.}
}

@article{yao,
    author = {Yao, Liang and Mao, Chengsheng and Luo, Yuan},
    date = {2019},
    title = {Graph Convolutional Networks for Text Classification},
    journal = {Proceedings of the AAAI Conference on Artificial Intelligence},
    volume = {33(01)},
    pages = {7370-7377},
    url = {https://doi.org/10.1609/aaai.v33i01.33017370}
}

@article{izmailov,
  doi = {10.48550/ARXIV.1803.05407},
  
  url = {https://arxiv.org/abs/1803.05407},
  
  author = {Izmailov, Pavel and Podoprikhin, Dmitrii and Garipov, Timur and Vetrov, Dmitry and Wilson, Andrew Gordon},
  
  title = {Averaging Weights Leads to Wider Optima and Better Generalization},
  
  publisher = {arXiv},
  
  year = {2018},
  
  copyright = {arXiv.org perpetual, non-exclusive license}
}

@article{bengio,
  doi = {10.48550/ARXIV.1206.5533},
  
  url = {https://arxiv.org/abs/1206.5533},
  
  author = {Bengio, Yoshua},
  
  title = {Practical recommendations for gradient-based training of deep architectures},
  
  publisher = {arXiv},
  
  year = {2012},
  
  copyright = {arXiv.org perpetual, non-exclusive license}
}

@misc{deepfindr,
  author = {DeepFindr},
  title = {A Graph Neural Network project on HIV data},
  year = {2021},
  publisher = {GitHub},
  journal = {GitHub repository},
  howpublished = {\url{https://github.com/deepfindr/gnn-project}},
}

@misc{philippe,
  author = {Philippe},
  title = {uvadlc notebooks},
  year = {2022},
  publisher = {GitHub},
  journal = {GitHub repository},
  howpublished = {\url{https://github.com/phlippe/uvadlc_notebooks/blob/master/docs/tutorial_notebooks/tutorial7/GNN_overview.ipynb}},
}

@software{pyg,
author = {Fey, Matthias and Lenssen, Jan Eric},
license = {MIT},
month = {5},
title = {{Fast Graph Representation Learning with PyTorch Geometric}},
howpublished = {\url{https://github.com/pyg-team/pytorch_geometric}},
year = {2019}
}

@incollection{py,
title = {PyTorch: An Imperative Style, High-Performance Deep Learning Library},
author = {Paszke, Adam and Gross, Sam and Massa, Francisco and Lerer, Adam and Bradbury, James and Chanan, Gregory and Killeen, Trevor and Lin, Zeming and Gimelshein, Natalia and Antiga, Luca and Desmaison, Alban and Kopf, Andreas and Yang, Edward and DeVito, Zachary and Raison, Martin and Tejani, Alykhan and Chilamkurthy, Sasank and Steiner, Benoit and Fang, Lu and Bai, Junjie and Chintala, Soumith},
booktitle = {Advances in Neural Information Processing Systems 32},
editor = {H. Wallach and H. Larochelle and A. Beygelzimer and F. d\textquotesingle Alch\'{e}-Buc and E. Fox and R. Garnett},
pages = {8024--8035},
year = {2019},
publisher = {Curran Associates, Inc.},
url = {http://papers.neurips.cc/paper/9015-pytorch-an-imperative-style-high-performance-deep-learning-library.pdf}
}

@software{pyl,
author = {Falcon, William and {The PyTorch Lightning team}},
doi = {10.5281/zenodo.3828935},
license = {Apache-2.0},
month = {3},
title = {{PyTorch Lightning}},
howpublished = {\url{https://github.com/PyTorchLightning/pytorch-lightning}},
version = {1.4},
year = {2019}
}

%%%%%%%%%%%%%%%%%%%%%%%%%%%%%%%%%%%%%%%%%%%%%%%%%%%%%%%%%%%%

%%%%%%%%%%%%%%%%%%%%%%%%%%%%%%%%%%%%%%%%%%%%%%%%%%%%%%%%%%%%

\newpage
\appendix
\section{Societal impacts}\label{apdx: soc imp}
This work has had two fields of development: one in the use of machine learning to further physics and the other to use physics to research machine learning. The development of physical theory has limited short term societal impact, since particle physics is many years from finding implementation in technology and engineering. This paper contributes towards the effectiveness of the graph attention layer but only in its application to Feynman diagrams and has its limitations. It does not suggest it could be easily substituted for unethical datasets and does not promote its use in immoral situations. The datasets used were self-constructed and did not use any sensitive information. Creating a trained network is also fast and not energy intensive with Google Colab's free GPUs being sufficient with 12.68GB RAM used to train the whole dataset, so there is little environmental impact.

\section{Implementation}\label{apdx: imple}

%Information about hyper parameters, data splits optimization algorithm, loss, attention layer, dataset encoding, and maybe some other things from the method section of my original project. Answers to checklist question
\subsection{Dataset encoding}\label{apdx: data enc}
Figure \ref{fig: graph encoding} gives an example Feynman diagram and a graphical depiction of its encoding to a graph.

\begin{figure}[t]
    \begin{center}
        \includegraphics[width=0.8\textwidth]{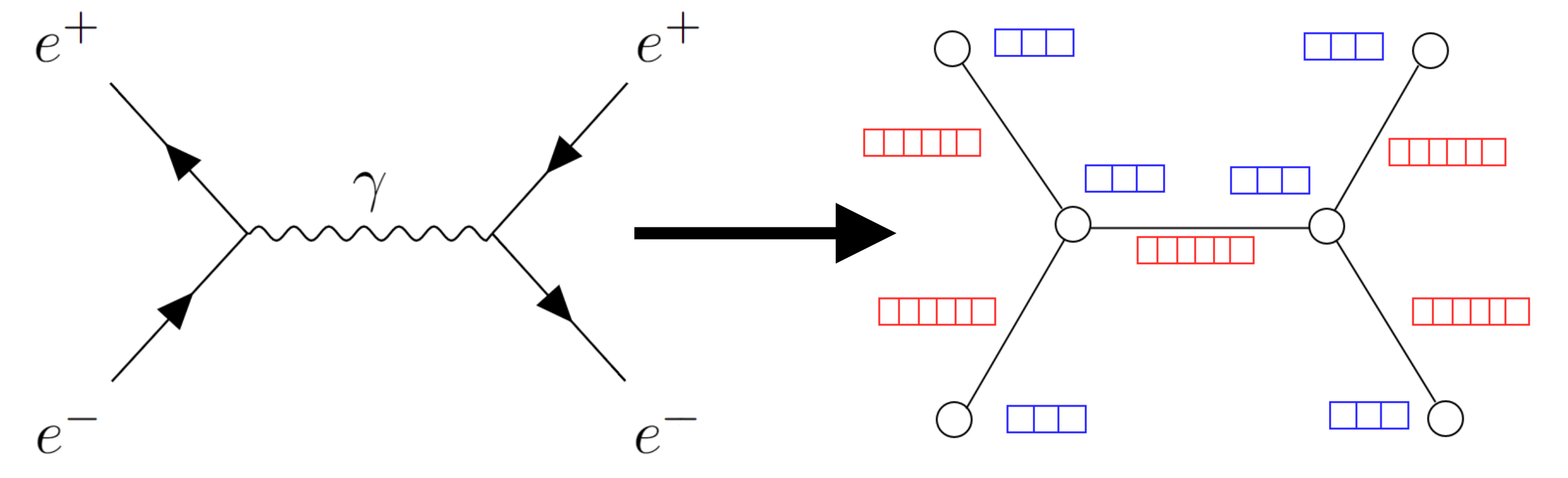}
    \end{center}\caption{Encoding a Feynman diagram as a graph. The blue vectors are the node features and the red are the edge features.}
    \label{fig: graph encoding}
\end{figure}

The nodes were also encoded as to make the graph orientable by capturing information about the time-like direction of the interaction. This was achieved by specifying which nodes are initial state particles, final state particles and virtual particles, using a one-hot encoding.

\begin{equation}
    \text{node vector} = [\text{initial},\text{virtual},\text{final}]
\end{equation}

\subsection{Training details}\label{apdx: train det}

The network was constructed in PyTorch \cite{py}, with the aid of PyTorch Geometric \cite{pyg} and PyTorch Lightning \cite{pyl}. We used the help of several GitHub repositories including the work by DeepFindr \cite{deepfindr} and the tutorial from Pytorch Geometric \cite{philippe}. All of the code was written in Google Colab using their free GPUs, with about 13GB of RAM total.

The loss was generated with a variety of loss functions including mean square error (MSE), absolute error ($L^1$), and LogCosh with the loss of a batch being calculated by the sum of the individual losses. Ultimately all losses performed similarly, but \(L^1\) loss was primarily used in this study as slight improvement on accuracy was found. A stochastic gradient descent optimizer with momentum 0.4 was used. Damping was also tested but did not improve the performance. The learning rate was started high at 1 but a scheduler that reduced the learning rate when the loss plateaued was implemented to hone in on the minima. The mini-batch size used depended on the dataset but was kept at powers of 2, between 16 and 64, to maximize storage use. This was a good compromise of speed and performance. The graph representation vector had dimension that depended on the complexity of the dataset, but reached a size of at most 8. This was sufficient for the network to encode the relevant information to pass to the FCN. The activation functions used throughout are ReLU and Leaky ReLU, these were fast to train and did not give stagnation problems that sigmoid and tanh do.

The number of dense neurons in the FCN was progressively increased if the model under-fitted, but a value between 64 and 128 was enough to be able to fit to the functions for this paper. The depth of the GNN was kept to only 2 or 3 layers, due to the global node.

\subsection{Runtime}\label{apdx: runtime}

The GNN training was completed on the order of minutes. Figure \ref{fig: e only 1 epoch} shows how the FCN begins to fit to the targets after just one epoch. Figure \ref{fig: val loss} shows how the training time increases as the complexity of the dataset grows. The blue line represents training on just one graph, which is why it converges on a minimum so quickly. The orange line represents the most complex dataset, and takes longer than the pink line to reach the same level of loss, being overtaken at around 5000 steps. The slow start is due to the stochastic nature of the optimization algorithm. Once it finds a minimum, the pink line converges more quickly. 

Training time is largely due to the number of parameters to fit. As the dataset gets more complex, more parameters are required and so the parameter space to search is larger. With the large learning rates used, it is possible that steep local minima are also missed.

The sudden jumps in each line is due to the learning rate scheduler. As the model starts to oscillate around the minima, the speed is dropped so that the model does not overshoot.

\begin{figure}[t]
    \centering
    \begin{subfigure}[h]{0.45\textwidth}
        \centering
        \includegraphics[width=0.9\linewidth]{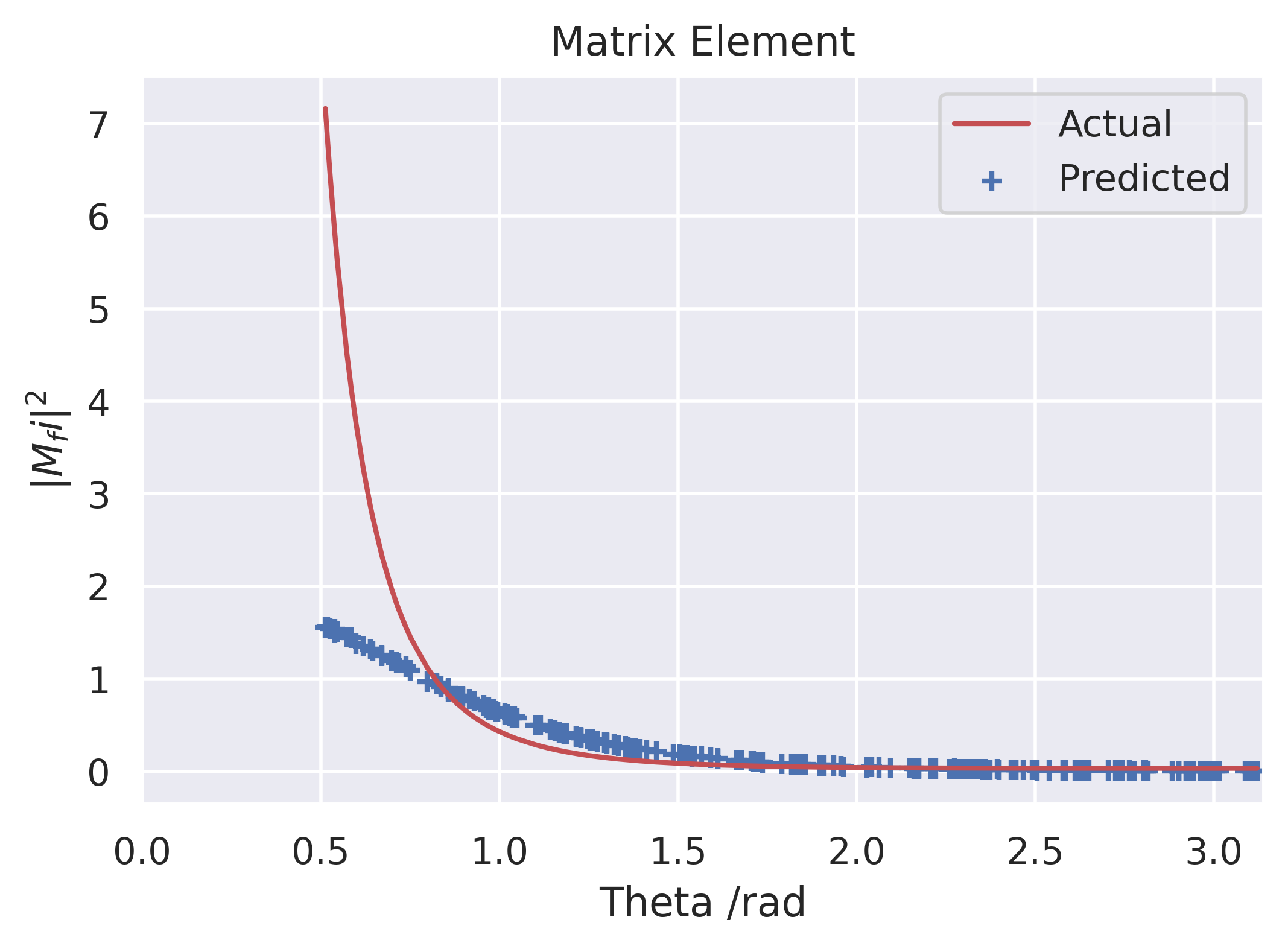}
        \caption{Training on \(e^-e^+\to e^-e^+\) Bhabha scattering diagrams after only one epoch of training}
        \label{fig: e only 1 epoch}
    \end{subfigure}
    \begin{subfigure}[h]{0.45\textwidth}
        \centering
        \includegraphics[width=0.9\linewidth]{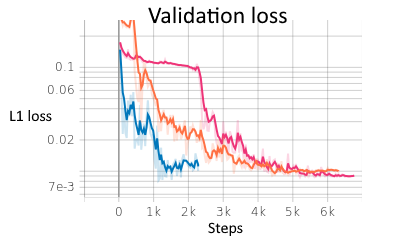}
        \caption{Graph showing the improvement of the network loss at the end of each epoch on the validation dataset. Blue: one graph. Pink: \(e^-e^+\to\mu^-\mu^+\) and \(e^-e^+\to e^-e^+\) graphs. Orange: whole dataset.}
        \label{fig: val loss}
    \end{subfigure}\caption{Graphs showing the speed to train the network}
\end{figure}

\subsection{Program code}\label{apdx: code}
 \url{https://github.com/Clearbloo/Feynman_GNN.git}

\newpage
\section{QCD}\label{apdx: QCD}

QCD was also trialed to test the models generalization to a different set of Feynman rules. The test process used was \(u\bar{u} \to t\bar{t}\) via a gluon s-channel. Since top quarks are massive, and not in chirality eigenstates, this also doubled as a test for the network to distinguish momentum dependence. There was a significant drop in performance, as shown in Figure \ref{fig: qcd} and Table \ref{tab: qcd}. However, the shape of the original QED predictions was retained and the model was able to identify the different curves based on momenta. This suggests the GNN can be scaled up to include more matrix elements and different sections of the standard model interaction Hamiltonian, with an increasingly large graph representation and improvements on implementation. This is expected as different Feynman rules need additional encodings on the graph representation.

\begin{table}[h]
    \caption{Loss and accuracy values for \(u\bar{u} \to t\bar{t}\) via gluon exchange}
    \centering 
    \begin{tabular}{lll}
        \toprule
        Test metric & Data \\
        \midrule
        Test accuracy (1dp) & 94.0\%\\
        Test accuracy (2dp) & 42.0\% \\
        Test accuracy (3dp) & 3.75\% \\
        Test L1 Loss & 0.0265\\
        \bottomrule
    \end{tabular}
    \label{tab: qcd}
\end{table}

\begin{figure}[h]
    \centering
    \includegraphics[width=0.8\textwidth]{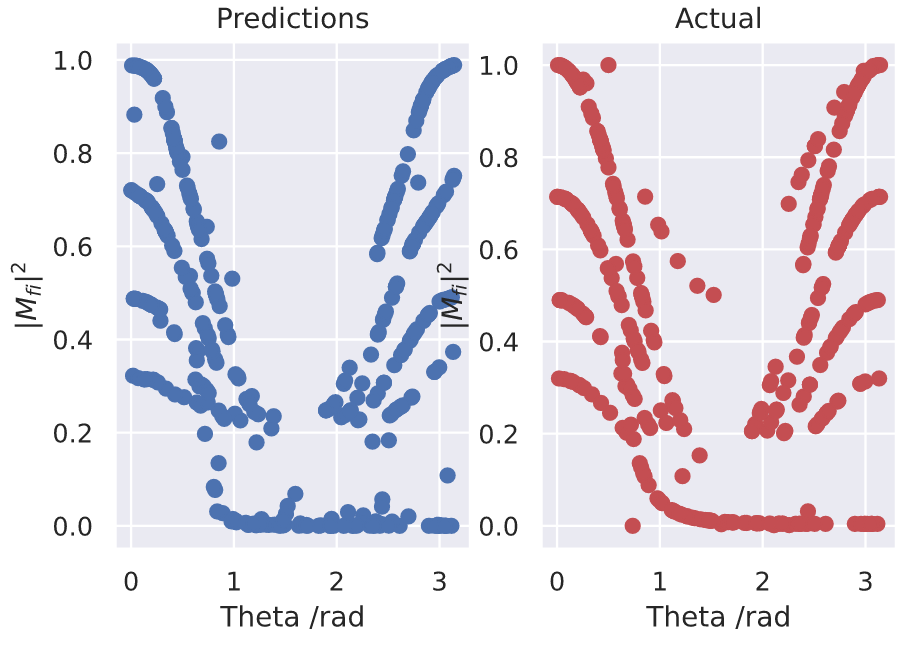}
    \caption{QCD results}
    \label{fig: qcd}
\end{figure}

%\section{Future Work}
%\begin{figure}[b]
%     \begin{center}
%         \includegraphics[width=0.5\textwidth]{figures/TGN.png}
%     \end{center}\caption{TGN}
%     \label{fig: TGN}
%\end{figure}

\end{document}